\documentclass[12pt,aps,prb,superscriptaddress,amsmath,amssymb,longbibliography]{revtex4-1}

\usepackage{graphicx,subfigure}% Include figure, and subfigures
\usepackage{dcolumn}% Align table columns on decimal point
\usepackage{bm}% bold math
\usepackage{epsf,color}
\usepackage{amsfonts}% Needed for \overbar, but did not work anyway
\usepackage{float, rotating, rotfloat}
\usepackage[section]{placeins} % Prevent figures from moving across sections.
\usepackage{gensymb} % for degree symbol
\usepackage{setspace}
\usepackage{amsmath}
\usepackage{physics} % for roman derivative, \dd and \dv

\usepackage{algcompatible, algpseudocode}
\usepackage[floatrow]{trivfloat}
\trivfloat{algorithm}

\usepackage{etoolbox}
\makeatletter
\preto{\@verbatim}{\topsep=0pt \partopsep=0pt }
\makeatother

\input Lectures.def

\def\beq{\begin{eqnarray}}
\def\eeq{\end{eqnarray}}

\begin{document}

\title{Energy needed to propel a tiny spacecraft to Proxima Centauri,
and,
an unstated assumption in Einstein's 1905 paper}

\author{C. J. Umrigar\footnote{CyrusUmrigar@cornell.edu}}
\affiliation{
Laboratory of Atomic and Solid State Physics,\\
Cornell University, Ithaca, NY 14853, USA.}

\author{Tyler A. Anderson\footnote{taa65@cornell.edu}}
\affiliation{
Laboratory of Atomic and Solid State Physics,\\
Cornell University, Ithaca, NY 14853, USA.}

\begin{abstract}
The Breakthrough Starshot project aims to send a tiny 2 gram spacecraft to Proxima Centauri
propelled by a light sail and powerful Earth-based lasers.
%The published estimates of the required
%laser energy are inaccurate because they do not properly account for the well known fact that
%almost all of the propulsion would occur during the very early stage of the flight.
We provide two derivations of the laser energy required to propel the spacecraft and give
the reader the opportunity to decide which one is correct before providing the answer.
In the second part of this paper we point out that one of the formulae in Einstein's amazing 1905 paper
is correct only in certain limits, but Einstein fails to mention that.  This has caused
some confusion in the Breakthrough Starshot literature.
\end{abstract}

\maketitle

\section{Introduction}

%We can select fontsize anywhere using a command like \fontsize{8mm}{9mm}\selectfont
%The rough mappings are
%\Large        ?
%\large        \fontsize{4.1mm}{4.9mm}\selectfont
%\normalsize   \fontsize{3.7mm}{4.5mm}\selectfont
%\small        \fontsize{3.4mm}{4.2mm}\selectfont
%\footnotesize \fontsize{3.0mm}{3.8mm}\selectfont
%\scriptsize   \fontsize{2.7mm}{3.5mm}\selectfont
%\tiny         \fontsize{2.2mm}{2.9mm}\selectfont

The prospects for space travel capture the human imagination and are the subject of much scientific
and engineering endeavor, as well as creative science fiction.
The two Voyager spacecraft, launched in 1977, are the fastest human-made objects currently travelling in interstellar space
at approximately 62,000 km/hour or $6 \times 10^{-5} c$, where $c$ is the speed of light in vacuum.
In Table~\ref{tab:distances} we show the distances of some well-known
%astronomical objects and time it would take light or a Voyager-like spacecraft to get there.
astronomical objects and the time it would take the Voyager spacecraft to get there.
%If one limits oneself to missions that can occur within a human lifetime, it is apparent that the entire
%solar system is accessible using current technology, but, the closest galaxy is not accessible even
If one limits oneself to missions that can occur within a human lifetime, it is apparent that while the entire
solar system is accessible using current technology, the closest galaxy is not accessible even
taking into account any reasonable projections of improvements to technology.
(One may be tempted to think that because of relativistic time-dilation, one could travel arbitrarily
far in a human lifetime, but in fact the energy requirements to approach the speed of light are enormous.)
%Travel to the closest star, Proxima Centauri, is in between -- it takes too long at Voyager speeds, but would take
Travel to the closest star, Proxima Centauri, is in-between -- it takes too long at Voyager speeds, but would take
only 21 years or 8.5 years of Earth-frame time if one could have spacecraft with a speed of 0.2 $c$ or 0.5 $c$
respectively.

\begin{table}[htbp]
\label{tab:distances}
\caption{Distances to various astronomical objects in light travel and Voyager travel units.}
\begin{tabular}{lll}
\hline
Object & Distance & Time on Voyager\\
\hline
Moon & 1.3 light seconds & 6 hours \\
Sun  & 8 light minutes & 93 days \\
Pluto  & 5.3 light hours & 10 years \\
\color{blue} Closest star (Proxima Centauri) & \color{blue} 4.24 light years & \color{blue} 71,000 years\\
%Milky Way size & $10^5$ light years\\
Closest galaxy (Canis Major Dwarf) & 25,000 light years & 420,000,000 years\\
\hline
\end{tabular}
\end{table}

The outline for the rest of the paper is as follows.  In Section~\ref{sec:rocket} we dispel any notion
that advances in chemical rocket technology could get us there.
In Section~\ref{sec:photon} we point out that photon rockets are not a viable solution because there
is no known technology that can convert a sufficiently large fraction of rest mass into photons.
In Section~\ref{sec:sail} we discuss the Breakthrough Starshot project, which aims to propel a tiny spacecraft weighing about 2 grams
using the pressure exerted by laser light on a light sail.
Major technical problems would have to be overcome for this project to succeed.
Instead, here we focus on the basic question of how much laser energy is needed because
that is a trickier question than first meets the eye.
We provide two derivations, which give
two different expressions for the required energy that differ greatly at relativistic speeds.
We encourage the reader to ponder which is the correct result and why, before reading the
explanation given at the end of the section.
In Section~\ref{sec:Einstein} we derive the expression for the energy of light reflected from
a moving perfect mirror and find that the expression derived by Einstein in his famous 1905
paper is correct only in limiting cases.
%In the supplementary material
Next, we discuss some confusion
in the Breakthrough Starshot literature related to this expression.

%{\color{red} Maybe we should also mention that Starshot is a fly-by mission because it would take far more fuel mass to deccelerate the rocket?}

\section{Chemical rockets are not the answer}
\label{sec:rocket}

The first question that may occur to the reader is whether improvements to conventional
chemical rocket technology could propel spacecraft to a good fraction of the speed of light.
The answer is a resounding no!

Chemical rockets eject hot gas, obtained by burning fuel, in one direction to propel the remainder
of the rocket in the opposite direction.
%The rocket equation relates $M$, the mass of the entire rocket $M$ (most of which is comprised of the fuel), to the mass of the payload $m$ (
The rocket equation relates
%$M$, the mass of the entire rocket, most of which is comprised of the fuel, to 
the mass of the entire rocket $M$ (most of which is comprised of the fuel), to 
%$m$ the mass of the payload (the part of the rocket that travels to the final destination),
the mass of the payload $m$ (the part of the rocket that travels to the final destination),
%$v_{\rm exh}$, the velocity of the exhaust gases, and,
the velocity of the exhaust gases $v_{\rm exh}$, and
%$v$, the desired eventual speed of the payload.
the desired eventual speed of the payload $v$.
A typical value for $v_{\rm exh}$ is 3 km/s = $10^{-5}\,c$.
Using $v=0.5\,c$, the rocket equation gives
\beq
M &=& m \; e^{v \over v_{\rm exh}} \;=\; m\; e^{0.5 c \over 10^{-5} c} \;=\; m\; e^{50000}
\;\approx\; m\; 10^{21715} \mbox{\color{black} \hskip 3mm (for $v=0.5 c$)}.
\eeq
This is truly an astronomically large mass, $M$, even for a tiny payload!
The reason is that early in the flight fuel is being burned to accelerate not just the payload but also the rest of the fuel.
In fact, the Voyager spacecrafts achieved even their rather modest speeds not by just
using chemical rocket propulsion, but by also using
gravity assists from Jupiter, Saturn and Uranus.

\section{Photon rockets are not the answer}
\label{sec:photon}

Since $M/m$ is exponential in ${v / v_{\rm exh}}$, an enormous reduction in the mass, $M$, can
be achieved by increasing $v_{\rm exh}$.  Let's carry this to the logical extreme, as has been proposed
before~\cite{FrenchBook}, and say that instead
of emitting exhaust gases, the rocket emits a collimated beam of photons.
Let $M_0$ be the initial rest mass of the rocket, $m_0$ be the rest mass of the payload and
$E_{\rm photon}$ be the energy of the emitted photons.
Then, from energy and momentum conservation we have
\beq
M_0 c^2 &=& \gamma m_0 c^2 + E_{\rm photon} \mbox{\color{black} \hskip 5.3mm (Energy cons.)} \\
\pvec_{\rm photon} &=& -\pvec_{\rm rocket} \mbox{\color{black} \hskip 20mm (Momentum cons.)}\\
\mbox{\color{black} so \hskip 2mm} p_{\rm photon} \;=\; p_{\rm rocket} &=& \gamma m_0 v \;\equiv\; \gamma m_0 c \beta
\eeq
Since the energy of photons is $c$ times their momentum, $E_{\rm photon}=\gamma m_0 c^2 \beta$.
Plugging this back into the energy conservation equation, we have
{\beq
M_0 {c^2} &=& \gamma m_0 {c^2} + \gamma m_0 {c^2} \beta \\ %\mbox{\color{black} \hskip 3mm (since $E_{\rm photon} = c \, p_{\rm photon}$)}\\
M_0 &=& \gamma(1+\beta) m_0 \;=\; \sqrt{1+\beta \over 1-\beta} m_0 \\
&\to&
\left\{\begin{array}{llll}
(1 + \beta) m_0, & \mbox{\color{black} when \hskip 2mm} \beta \to 0 \; (\gamma \to 1) \\
2 \gamma m_0, & \mbox{\color{black} when \hskip 2mm} \beta \to 1 \; (\gamma \gg 1) \\
\sqrt{3} m_0 = 1.73 m_0, & \mbox{\color{black} when \hskip 2mm} \beta \to 1/2
\end{array} \right.
\eeq}
This is an enormous improvement compared to a chemical rocket, but is still not feasible with
current technology for the following reason.
We have assumed that mass can be converted into a collimated beam of photons with 100\% efficiency
but there is no way to do this, aside from matter-antimatter annihilation.
There is no known way to make and store this quantity of antimatter.
Nuclear reactors are not the answer -- a fission reactor converts at most only $\sim 0.15$\% of its rest mass into heat,
and, further, the efficiency for converting heat into laser light is small.

\section{Light sails and the Breakthrough Starshot project}
\label{sec:sail}

Another idea is to leave the energy source that propels the spacecraft on Earth, or, in Earth orbit.
This can be achieved by using the pressure of laser light shining on a light sail to propel the spacecraft.
The goal of the Breakthrough Starshot project~\cite{BreakthroughStarshot} is to propel a tiny spacecraft
equipped with a light sail, weighing only about 2 grams, with an intense light beam from a phased array of lasers
to something like $0.2\,c$.
This project has \$100 million in seed funding from science philanthropist Yuri Milner, and
many distinguished scientists and engineers are associated with the project~\cite{BreakthroughStarshot}.
Needless to say, some very tough technical problems need to be overcome:
\begin{enumerate}
\itemsep 1em
\item Deploying a sail that is sufficiently light, with a sufficiently high area and a high reflectivity.
%It is acceptable if the transmission is not close to zero, but if the absorption is not close to zero the
%sail will vaporize.
%At minimum the absorptivity must be close to zero to avoid vaporizing the sail (but the
%transmissivity need not be very small).
It is helpful to have a reflectivity close to one, but it is essential that the absorptivity is
very close to zero so that the sail does not vaporize.
Further the sail must maintain its shape and orientation while it is being propelled.
\item Making the camera and the communication devices really tiny, so that the total rest mass of the
spacecraft is just 2 grams.
\item Having powerful enough lasers with a sufficiently small angular spread.
\end{enumerate}

Here, we do not concern ourselves with these technical problems or the likelihood that they will
ever be surmounted.  Instead we address the basic question of how much laser light energy
is needed to propel the spacecraft to velocity $v$.

We use standard relativistic notation, namely,
\beq
\beta &=& {v \over c}, \\
\gamma &=& {1 \over \sqrt{1-\beta^2}}.
\eeq

We now give 2 derivations that give different expressions for the energy
needed and then give the reader the opportunity to think about which is right and relevant, which is
wrong or not relevant, and why.

\subsection{Derivation 1}

We find the energy, $E_I$, of the laser light needed to accelerate the spaceship with rest mass $m_0$
from an initial velocity $\beta_i$ to a final velocity $\beta_f$.
%Denote energy of reflected light by $E_R$.
Denote the energy of the reflected light by $E_R$.
\beq
\label{eq:energy_cons}
\gamma_i m_0 c^2 + E_I &=& \gamma_f m_0 c^2 + E_R      \mbox{\color{black} \hskip 9mm (energy cons.)} \\
\label{eq:momentum_cons}
\gamma_i m_0 c \beta_i + {E_I \over c} &=& \gamma_f m_0 c \beta_f - {E_R \over c}  \mbox{\color{black} \hskip 6mm (momentum cons.)}
%\mbox{\color{black} i.e., \hskip 9mm} E_I  &=& \gamma \beta m_0 c^2 - E_R \mbox{\color{black} \hskip 5mm (momentum cons.)}
\eeq
Adding Eq.~\ref{eq:energy_cons} to $c$ times Eq.~\ref{eq:momentum_cons}, we eliminate $E_R$ and get
\beq
\gamma_i(1+\beta_i)m_0 c^2 + 2E_I &=& \gamma_f (1+\beta_f) m_0 c^2 \nonumber \\
\mbox{\color{black} i.e., \hskip 5mm} E_I  &=& {1 \over 2} \left(\gamma_f \left(1+\beta_f\right)-\gamma_i \left(1+\beta_i\right)\right) m_0 c^2 \nonumber \\
&=& \boxed{ {1 \over 2} \left(\sqrt{1+\beta_f \over 1-\beta_f} -\sqrt{1+\beta_i \over 1-\beta_i} \right) m_0 c^2 }
\label{eq:energy_right_beta_i}
%&=& \left({2 \over \sqrt{3}} \left({3 \over 2}\right)-1\right) {m_0 c^2 \over 2}\mbox{\color{black} \hskip 5mm (for $\beta={1 \over 2})$}\\
%&=& {1 \over 2} \left({\sqrt{3/2} \over \sqrt{1/2}} -1\right) {m_0 c^2 \over 2}\mbox{\color{black} \hskip 5mm (for $\beta={1 \over 2})$}\\
%&=& {\left(\sqrt{3}-1\right) \over 2} m_0 c^2 \;\approx\; 0.37 m_0 c^2 \mbox{\color{black} \hskip 5mm (for $\beta={1 \over 2})$}
\eeq
If the initial velocity, $\beta_i=0$, then
\beq
E_I  &=& {1 \over 2} \left(\sqrt{1+\beta_f \over 1-\beta_f} -1 \right) m_0 c^2
\to \left\{\begin{array}{lll}
 {\beta_f \over 2} m_0 c^2, \mbox{\color{black} \hskip 34mm when \hskip 2mm} \beta_f \to 0\\[2mm]
 {\left(\sqrt{3}-1\right) \over 2} m_0 c^2 \;\approx\; 0.37 m_0 c^2, \mbox{\color{black} \hskip 2mm when \hskip 1.3mm} \beta_f={1 \over 2}\\[2mm]
 \left(\gamma_f -{1 \over 2}\right) m_0 c^2, \mbox{\color{black} \hskip 22mm when \hskip 2mm} \gamma_f \gg 1 \; (\beta_f \to 1).
\end{array} \right.
\label{eq:energy_right}
\eeq
Note that the final velocity, $\beta_f$, depends only on the total incident energy, $E_I$,
and not on the energy schedule.

At small $\beta_f$, the required energy reduces to the nonrelativistic expression,
\beq
E_I &=&  {\beta_f \over 2} m_0 c^2,
\eeq
as it must.
That this is the nonrelativistic expression can be seen as follows.
%$m_0v$ is the momentum and it must equal twice the momentum of the incident light,
%which is $2E/c$.  So, $E_I = {c \over 2} m_0 v = {\beta\over 2} m_0 c^2$.
In the small $\beta_f$ limit, the reflected light momentum, $- {E_R \over c}$, is equal in magnitude to
the incident light momentum, ${E_I \over c}$, so that the final momentum of the craft, $m_0 v$, must equal $2E_I/c$.
So, $E_I = {c \over 2} m_0 v = {\beta_f \over 2} m_0 c^2$.

As an aside, we note that, instead of solving for $E_I$ as a function of $\beta_i$ and $\beta_f$,
as in Eq.~\ref{eq:energy_right_beta_i}, we can solve for $\beta_f$ as a function of $\beta_i$ and $E_I$ to obtain
\beq
\beta_f &=& {\beta_i + 2 \sqrt{1-\beta_i^2} \,r + 2 (1-\beta_i) r^2 \over
                   1 + 2 \sqrt{1-\beta_i^2} \,r + 2 (1-\beta_i) r^2},
\label{eq:beta_f}
\eeq
where $r=E_I/(m_0c^2)$.
Eq.~\ref{eq:energy_right} is the same as Eq.~17 of Ref.~\onlinecite{Kip-AJ-17} but our derivation is
simpler.
Eq.~\ref{eq:beta_f} is much simpler than Eq.~3 of Ref.~\onlinecite{Kip-AJ-17}, but they are equivalent
as can be checked by plugging in arbitrary numerical values for $\beta_i$ and $r$.

%\centerline{\includegraphics[width=1.\textwidth]{StarShot.eps}}

%\centerline{\Huge \color{red} \bf Derivation 2:}
\subsection{Derivation 2}

We now give the second derivation, which consists of the following steps:
\begin{enumerate}
%\bf
\item Relate the laser power to the force on the spacecraft.
\item Relate the force to the acceleration.
\item Integrate over time to relate the laser energy to the spacecraft velocity.
\end{enumerate}

\noindent \underline{\bf 1) Relate the laser power to the force on the spacecraft}\\
If photons are emitted at intervals of $T$, the photons are spaced $cT$ apart in the Earth frame.
In the Earth frame photons are moving with speed $c-v$ relative to the spacecraft, so in the Earth
frame the photons are received by the spacecraft
\vskip -6mm
\beq
\label{eq:Tr}
T_r &=& {cT \over (c-v)} \;=\; {T \over 1-\beta}
\eeq
apart in time.

%Now, in the Earth frame, the spacecraft receives photons of the same energy, $\nu_I$, that the laser emits, but
Now, in the Earth frame, the spacecraft receives photons of the same energy that the laser emits, but
the reflected photons have a lower energy.  To figure that out, following the argument in Einstein's
1905 paper~\cite{Ein-AdP-05b}, transform to the spacecraft frame and back.
In the spacecraft frame the energy of the received photons is Doppler shifted to
\beq
%E_R &=& E_I \sqrt{1-\beta \over 1+\beta}
E_I' &=& E_I \sqrt{1-\beta \over 1+\beta}
\label{eq:Doppler1}
\eeq
and the reflected photons have very nearly the same energy because the spacecraft mass is about $10^{33}$ times greater than the
relativistic mass of an optical photon. (The rest mass of the photon is of course zero, but the relativistic
mass is not.)
In the Earth frame, these reflected photons will be further Doppler shifted, so the energy in the Earth frame of
the reflected photons is
\beq
\label{eq:nu_refl}
%E_{\rm refl} &=& E_I \left({1-\beta \over 1+\beta}\right)
E_R = E_I' \sqrt{1-\beta \over 1+\beta} &=& E_I \left({1-\beta \over 1+\beta}\right)
\label{eq:Doppler2}
\eeq
%Now, the force on the spacecraft equals the rate of change of its momentum, i.e., $\displaystyle F={\mathrm{d} p \over \mathrm{d} T}$, the momentum of a photon is its energy divided by $c$, and, the laser power is the rate at
%Now, the force on the spacecraft equals the rate of change of its momentum, i.e., $\displaystyle F={\mathrm{d} p \over \mathrm{d} T}$, the magnitude of the momentum of a photon is its energy divided by $c$, and, the laser power is the rate at
Now, the force on the spacecraft equals the rate of change of its momentum, i.e., $\displaystyle F={\mathrm{d} p \over \mathrm{d} T} = {\Delta p \over T_r} $, the magnitude of the momentum of a photon is its energy divided by $c$, and, the laser power is the rate at
%which it is emitting energy, i.e., $\displaystyle P = {\mathrm{d} E_I \over \mathrm{d} T}$.
which it is emitting energy, i.e., $\displaystyle P = {\mathrm{d} E_I \over \mathrm{d} T} = {E_I \over T}$.
Hence, using Eqs.~\ref{eq:Tr} and \ref{eq:nu_refl}, we get
%Power of the laser light, $\displaystyle P = {\mathrm{d} E_I \over \mathrm{d} T}$, and its $\displaystyle F={\mathrm{d} p \over \mathrm{d} T}$, is (using Eqs.~\ref{eq:Tr} and \ref{eq:nu_refl})
\beq
%F \;=\; {E_I + E_{\rm refl} \over c T_r}  &=& {E_I \over c T} (1-\beta) \left (1 + {1-\beta \over 1+\beta} \right) \;=\; {P \over c} (1-\beta) \left({2 \over 1+\beta}\right) \nonumber\\
%F \;=\; {\Delta p \over T_r} \;=\; {E_I + E_{\rm refl} \over c T_r}
F \;=\; {\Delta p \over T_r} \;=\; {E_I + E_R \over c T_r}
%F \;=\; {\Delta p \over T_r} \;=\; {{E_I \over c} - {- E_R \over c} \over T_r}
&=& {1 \over c} {1-\beta \over T} E_I \left(1+{1-\beta \over 1+\beta}\right)
\;=\; {2 \over c} \left({1-\beta \over 1+\beta}\right) P \nonumber \\
\mbox{\color{black} i.e.} \quad \boxed{P = {c \over 2} \left(1+\beta \over 1-\beta \right)F}
&& \mbox{\hskip 2mm \color{black} \footnotesize (relate force to power)}
\label{eq:F_to_P}
\eeq
%(This agrees with the spacecraft frame analysis for a reflecting sail, i.e., $F=F'$ on previous slide.)

%\end{frame}
%----------------------------------------------------------------------------
%\begin{frame}[allowframebreaks=1,shrink,label=include1]\frametitle{Light Energy Needed to accelerate from 0 to $\bm \beta$}

\noindent \underline{\bf 2) Relate the force to the acceleration}\\
Next, we derive the relativistic expression relating force and acceleration.
\beq
%F \;=\; {2 P \over c}\left({1-\beta \over 1+\beta}\right)
F \;=\; {d (\gamma m_0 v) \over dt}
&=& {d (\gamma m_0 c \beta) \over dt} \;=\; m_0 c \left( {d (\gamma \beta) \over dt}\right) \nonumber \\
&=& m_0 c \left( {d \gamma \over d \beta}{d \beta \over d t} \beta + \gamma {d\beta \over dt} \right)
\;=\; m_0 c \left( {d \gamma \over d \beta} \beta + \gamma\right) {d\beta \over dt} \nonumber \\
&=& m_0 c \gamma \left(\beta^2\gamma^2+1\right) {d\beta \over dt}
\;=\; m_0 c\gamma (\gamma^2) {d\beta \over dt} \nonumber \\
%\;=\; m_0 c\gamma (\gamma^2) {d\beta \over dt} \mbox{\color{black} \tiny \hskip 3mm (${d\gamma \over d \beta} = \beta \gamma^3$, $\beta^2\gamma^2+1=\gamma^2$)} \nonumber \\
&=& \gamma^3 m_0 c {d\beta \over dt}.
%\,=\, \gamma^3 m_0 {dv \over dt}
\label{eq:F}
\eeq
The relativistic force is $\gamma^3$ times larger than the nonrelativistic expression.

\noindent \underline{\bf 3) Integrate over time to relate the laser energy to the spacecraft velocity}\\
%\beq
%E_I &=& \left({(2-\beta)\sqrt{1-\beta^2} \over 3(1-\beta)^2} -{2 \over 3} \right) {m_0 c^2 \over 2}
%\eeq
Using Eqs.~\ref{eq:F_to_P} and \ref{eq:F}:
\beq
E_I &=& \int_0^T P dt \,=\,
\int_0^T {c \over 2} \left(1+\beta \over 1-\beta \right) \gamma^3 m_0 c {d\beta \over {dt}} {dt} =
{m_0 c^2 \over 2} \int_0^{\beta_f} d\beta \left({1+\beta \over 1-\beta}\right) {1 \over (1-\beta^2)^{3/2}} \nonumber \\
&=& \boxed{\left({(2-\beta_f)\sqrt{1-\beta_f^2} \over 3(1-\beta_f)^2} -{2 \over 3} \right) {m_0 c^2 \over 2}}
%\to {m_0 c^2 \over 2} \beta \mbox{\color{black} \hskip 2mm when \hskip 2mm} \beta \to 0 \nonumber
%\to
%\left\{\begin{array}{lll}
%\beta {m_0 c^2 \over 2}, \mbox{\color{black} \hskip 6mm when \hskip 2mm} \beta \to 0 \\[2mm]
%{4 \over 3} \gamma^3 m_0 c^2, \mbox{\color{black} \hskip 2mm when \hskip 2mm} \gamma \gg 1
%\end{array} \right. \nonumber
\label{eq:energy_wrong}
\eeq

%\begin{equation}
%\boxed{E_I
%\;=\; \left({(2-\beta)\sqrt{1-\beta^2} \over 3(1-\beta)^2} -{2 \over 3} \right) {m_0 c^2 \over 2} }
%%\to {m_0 c^2 \over 2} \beta \mbox{\color{black} \hskip 2mm when \hskip 2mm} \beta \to 0 \nonumber
%\to
%\left\{\begin{array}{lll}
%{\beta \over 2} m_0 c^2, \mbox{\color{black} \hskip 6mm when \hskip 2mm} \beta \to 0 \\[2mm]
%{4 \over 3} \gamma^3 m_0 c^2, \mbox{\color{black} \hskip 2mm when \hskip 2mm} \gamma \gg 1
%\end{array} \right. \nonumber
%%E_I &=& {m_0 c^2 \over 6} \left({(2-\beta)\sqrt{1-\beta^2} \over (1-\beta)^2} - 2 \right)
%%\to {m_0 c^2 \over 2} \beta \mbox{\color{black} \hskip 2mm when \hskip 2mm} \beta \to 0
%\end{equation}

If we divide Eq.~\ref{eq:energy_wrong} by the power of the laser, we obtain an expression
for the time as a function of $\beta_f$ that has also been published several times
in the literature~\cite{Lub-ARX-16,KulLubZha-AJ-18,LubHet-AF-20,LubinBook} (see e.g. Eq. 11 of
Ref.~\onlinecite{KulLubZha-AJ-18}).

\subsection{Summary of results from derivations 1 and 2}
We have provided two derivations of the laser energy, $E_I$, needed to accelerate the spacecraft
from zero velocity to $\beta$ (we now drop the subscript, $f$, to simplify the notation),
and arrived at two different expressions (Eqs.~\ref{eq:energy_right} and \ref{eq:energy_wrong}),
which we repeat below for easy comparison:\\
%both of which are in the literature, as we mentioned before.
%We repeat these expressions below for easy comparison:\\
\underline{\bf Derivation 1 Result:}\\
\beq
\boxed{E_I \;=\;
\left(\sqrt{1+\beta \over 1-\beta} -1 \right) {m_0 c^2 \over 2}}
%&=& \left({2 \over \sqrt{3}} \left({3 \over 2}\right)-1\right) {m_0 c^2 \over 2}\mbox{\color{black} \hskip 5mm (for $\beta={1 \over 2})$}\\
%&=& {1 \over 2} \left({\sqrt{3/2} \over \sqrt{1/2}} -1\right) {m_0 c^2 \over 2}\mbox{\color{black} \hskip 5mm (for $\beta={1 \over 2})$}\\
%&=& {\left(\sqrt{3}-1\right) \over 2} m_0 c^2 \;\approx\; 0.37 m_0 c^2 \mbox{\color{black} \hskip 5mm (for $\beta={1 \over 2})$}
\;\to\; \left\{\begin{array}{lll}
 {\beta \over 2} m_0 c^2, \mbox{\color{black} \hskip 16mm when \hskip 2mm} \beta \to 0\\[2mm]
%{\left(\sqrt{3}-1\right) \over 2} m_0 c^2 \;\approx\; 0.37 m_0 c^2, \mbox{\color{black} \hskip 2mm when \hskip 1.3mm} \beta={1 \over 2}\\[2mm]
 \left(\gamma -{1 \over 2}\right) m_0 c^2, \mbox{\color{black} \hskip 6mm when \hskip 2mm} \gamma \gg 1 \; (\beta \to 1)
\end{array} \right.
\label{eq:energy_rightb}
\eeq

\underline{\bf Derivation 2 Result:}\\
\beq
\boxed{E_I
\;=\; \left({(2-\beta)\sqrt{1-\beta^2} \over 3(1-\beta)^2} -{2 \over 3} \right)  {m_0 c^2 \over 2}}
%\to {m_0 c^2 \over 2} \beta \mbox{\color{black} \hskip 2mm when \hskip 2mm} \beta \to 0 \nonumber
\to
\left\{\begin{array}{lll}
{\beta \over 2} m_0 c^2, \mbox{\color{black} \hskip 5mm when \hskip 2mm} \beta \to 0 \\[2mm]
{4 \over 3} \gamma^3 m_0 c^2, \mbox{\color{black} \hskip 1.5mm when \hskip 2mm} \gamma \gg 1 \; (\beta \to 1)
\end{array} \right.
%E_I &=& {m_0 c^2 \over 6} \left({(2-\beta)\sqrt{1-\beta^2} \over (1-\beta)^2} - 2 \right)
%\to {m_0 c^2 \over 2} \beta \mbox{\color{black} \hskip 2mm when \hskip 2mm} \beta \to 0
\label{eq:energy_wrongb}
\eeq
We encourage the reader to pause and think about which is correct and relevant, and which is incorrect or not
relevant.  We note that both have the correct nonrelativistic limit, so this cannot be the basis
for deciding.  On the other hand, in the highly relativistic limit ($\gamma \to \infty$) they
are dramatically different -- the first scales as $\gamma$, whereas the second scales as $\gamma^3$.

\subsection{Resolution of the seeming contradiction}
\label{sec:resolution}

The two expressions answer two different questions.  Eq.~\ref{eq:energy_rightb} answers the question:
%what is the velocity after energy $E_I$ has been received by the spacecraft.
how much energy, $E_I$, has been received by the spacecraft when its velocity is $\beta$.
Eq.~\ref{eq:energy_wrongb} answers the question:
how much energy, $E_I$, has been emitted by the laser when the velocity of the spacecraft is $\beta$,
assuming that the laser emits with constant power.
As the spacecraft moves away from Earth, a progressively larger fraction of the light emitted
by the laser is in transit between Earth and the spacecraft and that light has not yet imparted any
momentum to the spacecraft.  This is why the energy in Eq.~\ref{eq:energy_wrongb} is so much larger
than that in Eq.~\ref{eq:energy_rightb} for large $\gamma$.

\begin{figure}
\label{fig:energy}
\centerline{\includegraphics[width=.9\textwidth]{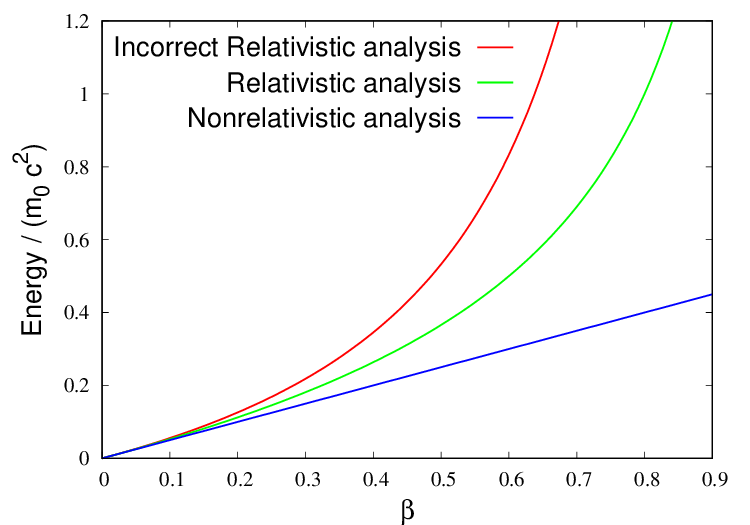}}
\caption{Comparison of the energy expressions in Eq.~\ref{eq:energy_rightb} and \ref{eq:energy_wrongb}
and the nonrelativistic expression for the energy needed to achieve velocity $\beta$.
%The incorrect analysis, used by the Breakthrough Starshot project, gives an energy that is 45\% too large at $\beta=0.5$ and 4.6 times too large at $\beta=0.9$.
}
\end{figure}

Which expression is relevant to the actual experiment?  The laser beam has a small but nonzero angular spread, $\theta$,
so part of the beam will miss the light sail once the spacecraft is at a distance $d > {L \over \theta} $,
where $L$ is the linear dimension of the light sail.  Hence it makes sense to shine the laser
during only the very early part of the mission, so that all of the laser light is captured
by the sail.
\footnote{Continuing to shine the laser beyond this point is wasteful because, for typical parameters, it results
in only about a 10\% increase in the terminal velocity of the spacecraft.}
The spacecraft will reach its terminal velocity once this energy reaches it.
Hence it is the first expression, Eq.~\ref{eq:energy_rightb}, that is the relevant one, and
that is fortunate since it gives the smaller energy estimate.
Eq.~\ref{eq:energy_wrongb} gives an energy that is 45\% too large at $\beta=0.5$, and 4.6 times too large at $\beta=0.9$.
Fig.~\ref{fig:energy} compares the energies of Eqs.~\ref{eq:energy_rightb} and \ref{eq:energy_wrongb} as
well as the nonrelativistic expression.
%Surprisingly, publications related to the Breakthrough Starshot project~\cite{Lub-ARX-16,KulLubZha-AJ-18,LubHet-AF-20,LubinBook} have repeatedly
%presented Eq.~\ref{eq:energy_wrongb} even though the authors are clearly aware that most of
%the energy in that expression has not reached the spacecraft, and moreover they have argued against~\cite{KulLubZha-AJ-18}
%the correct expression of Eq.~\ref{eq:energy_rightb} published (with needless caveats) in Ref.~\onlinecite{Kip-AJ-17}.

\subsection{Energy estimate for $\beta=0.5$}

We now estimate the energy needed to accelerate a spacecraft, with rest mass $m_0=2$ grams,
to $\beta=0.5$.
From Eq.~\ref{eq:energy_rightb},
\beq
E_I &=& {1 \over 2} \left(\sqrt{1+\beta \over 1-\beta} -1 \right) m_0 c^2 \nonumber \\
&=& 0.37 \times 0.002 \, {\rm kg} \times (3 \times 10^8 {\rm m/s})^2 \;=\; 7 \times 10^{13} {\rm J}
%&=& {7 \times 10^{13} {\rm J} \over 3600 \;{\rm J / W\,hour}}
%\;=\; 1.9 \times 10^{10} \,{\rm W\,hours}
\;=\; 19 \;{\rm Gigawatt\,hours}
\eeq

A typical power plant produces 1 Gigawatt of power.  Assuming that the propulsion
occurs during the first hour of flight, this would require harnessing the power
of 19 power plants, but only for an hour.  Since conversion of electric power to
laser light is not 100\% efficient, in practice more energy would be required.

\section{Energy of light reflected by a moving perfect mirror}
\label{sec:Einstein}

We now find the expression for the energy reflected back by a moving perfect mirror.
Again, let $E_I$ and $E_R$ be the energy of incident and the reflected light respectively,
$m_0$ be the rest mass of the mirror, $\beta_i$ and $\beta_f$ be its initial and
final velocities, and, $\nhat$ and $-\nhat$ be the direction of the incident and reflected light.
We again use conservation of energy and momentum, but use of the 4-vector formalism
simplifies the derivation.  Equating 4-momenta before and after:
\beq
{E_I \over c}(1,\nhat) + \gamma_i m_0 c\left(1, \beta_i \nhat\right) &=& {E_R \over c}(1,-\nhat) + \gamma_f m_0 c\left(1, \beta_f \nhat\right) \\
\mbox{\color{black} i.e., \hskip 4mm} {E_I \over c}(1,\nhat) - {E_R \over c}(1,-\nhat) + \gamma_i m_0 c\left(1, \beta_i \nhat\right) &=& \gamma_f m_0 c\left(1, \beta_f \nhat\right)
\eeq
Squaring the 4-momenta we obtain,
\beq
{(m_0c)^2} -4 {E_I E_R \over c^2} +2 E_I\gamma_i m_0 (1-\beta_i) -2 E_R\gamma_i m_0 (1+\beta_i) &=& {(m_0c)^2} \\
\mbox{\color{black} i.e., \hskip 4mm} \left((1+\beta_i)\gamma_i + 2 {E_I \over m_0 c^2}\right) E_R &=& (1-\beta_i) \gamma_i E_I
\eeq
\vskip -15mm
\beq
\mbox{\color{black} i.e., \hskip 4mm} \boxed{E_R \;=\; {(1-\beta_i) \over 1+\beta_i +2 r/\gamma_i} E_I}, \mbox{\hskip 3mm where \hskip 4mm} r = {E_I \over m_0 c^2}
\label{eq:energy_refl}
%\left( 2 {E_I \over c^2} + (1+\beta_i2 (E_I-E_R)\gamma_i m_0 (1-\beta_i) &=& \Cancel[red]{(m_0c)^2}
\eeq
%\small
%where $r = {E_I \over m_0 c^2}$, which reduces to Einstein's expression if $m_0 = \infty$,
%but he fails to state the assumption.
Note that the frequency of the reflected photons changes over time as the mirror speeds up;
the reflected energy, $E_R$, is the total reflected energy after energy $E_I$ has been shone on
the mirror which had initial velocity $\beta_i$.
For a single photon and an electron as the target, this is the Compton scattering formula, except
that we allow the initial velocity, $\beta_i$, to be nonzero, and restrict the scattering angle
to be 180$\degree$.

Einstein, in Section 8 of his first paper on special relativity~\cite{Ein-AdP-05b} in 1905, uses the argument
presented in the section leading up to Eq.~\ref{eq:Doppler2} to derive the expression
\beq
\nu_R \;=\; {1-\beta_i \over 1+\beta_i} \nu_I.
\label{eq:energy_refl_Einstein}
\eeq
The more general expression of Eq.~\ref{eq:energy_refl} reduces to Eq.~\ref{eq:energy_refl_Einstein} when either $m_0 \to \infty$,
or, if $\beta_i$ is interpreted as the instantaneous velocity and one considers classical
electrodynamics so that the energy of photons is not quantized and the energy received, $E_I$,
can be infinitesimal.  On the other hand for
quantized photons, Eq.~\ref{eq:energy_refl_Einstein} is never exactly correct though the
correction term, $r$, is $10^{-33}$ for optical photons and a 2 gram spacecraft.
Interestingly, just three months prior to his special relativity paper, Einstein had published
his paper on the photoelectric effect~\cite{Ein-AdP-05a} (for which he was awarded the Nobel prize); nevertheless
in his paper on special relativity he considers classical electrodynamics only.

Did Einstein realize that he was making an assumption when he derived Eq~\ref{eq:energy_refl_Einstein}?  Surprisingly, it appears he did not.  He is rather emphatic in his paper:\\
\begin{quote} \color{blue} \it
``All problems in the optics of moving bodies can be solved by the method
here employed. What is essential is, that the electric and magnetic force of the
light which is influenced by a moving body, be transformed into a system of
co-ordinates at rest relatively to the body. By this means all problems in the
optics of moving bodies will be reduced to a series of problems in the optics of
stationary bodies."
\end{quote}

%{\color{red} Maybe add:
%We emphasize that, although Einstein's statement is correct and Eq.~\ref{energy_refl_Einstein} can indeed be employed to correctly solve the problem of light reflecting from a moving mirror, it is directly applicable only when the photon energy $E_I$ is much smaller than the rest energy of the mirror $m_0 c^2$ or $r = {E_I \over m_0 c^2} \ll 1$.
%Otherwise, Eq.~\ref{energy_refl_Einstein} may be used by splitting the total photon energy into infinitesimal parts and integrating.
%}

\subsection{The Breakthrough Starshot Project and Einstein's Formula}

%The failure of Einstein to state his assumptions has caused some confusion in the Breakthrough Starshot literature.
%Kipping derived Eq.~\ref{eq:energy_right} (though in a needlessly complicated way with unnecessary caveats)
%but incorrectly ascribed the difference between Eq.~\ref{eq:energy_right} and Lubin and coworker's derivation
%of Eq.~\ref{eq:energy_wrong} to their use of Einstein's Eq.~\ref{eq:energy_refl_Einstein} instead of
%Eq.~\ref{eq:energy_refl}.  In fact Lubin used Eq.~\ref{eq:energy_refl_Einstein} correctly because
%$\beta_i$ in his derivation is the instantaneous value of $\beta$ and he integrates $\beta$ over time to
%arrive at the final $\beta$.  Rather, the ``error" in Lubin et al.'s derivation is that they are not
%calculating energy relevant to the experimental situation, as explained in Sec.~\ref{sec:sail}.

There has been some confusion in the Breakthrough Starshot literature, in part caused by Einstein's
failure to state his assumptions when deriving Eq.~\ref{eq:energy_refl_Einstein}.
Kipping in his paper~\cite{Kip-AJ-17} attributed the large (several percent) larger time that
Lubin and coworkers found for the time required for the spacecraft to reach a given velocity
to Lubin et al. using Eq.~\ref{eq:energy_refl_Einstein} rather than Eq.~\ref{eq:energy_refl}.
In reality, that difference is due to Lubin computing a quantity that is not very relevant
to the actual space mission, as explained in Section~\ref{sec:resolution}.
Later, Kipping published an erratum~\cite{Kip-AJ-17-err}
in which he computed the same quantity as Lubin and found a much smaller difference from Lubin et al.,
which he again attributed to Lubin et al. using Eq.~\ref{eq:energy_refl_Einstein} rather than Eq.~\ref{eq:energy_refl}.
However, even this much smaller difference is much larger than the actual difference, which is
of ${\cal O}(r) \approx 10^{-33}$ as shown in the Supplementary Material.
It is worth noting that the difference between Eq.~\ref{eq:energy_refl} and Eq.~\ref{eq:energy_refl_Einstein}
is large if the energy $E_I$ incident on the sail since the time that the sail had velocity $\beta_i$
is sufficiently large that $r=E_I/(m_0 c^2)$ is ${\cal O}(1)$.  On the other hand, in our Derivation 2,
which is equivalent to that of Lubin et al., $\beta$ is the instantaneous velocity that is integrated
over as it changes, and the difference between using
%classical E\&M expression Eq.~\ref{eq:energy_refl_Einstein}
%and  Eq.~\ref{eq:energy_refl}, which is exact for quantized photons,
Eq.~\ref{eq:energy_refl_Einstein} (the classical electrodynamics expression)
and Eq.~\ref{eq:energy_refl} (which is exact even for quantized photons)
is ${\cal O}(r)$, where $r=h\nu/(m_0 c^2) \approx 10^{-33}$ for optical phonons and $m_0 \approx 2$ gram.

\section{Conclusions}

In the first portion of this paper, we discussed the laser energy needed to propel a tiny spacecraft to Proxima Centauri.  We derived
two formulae and explained why the one most commonly presented in the literature is not relevant to the problem.
In the second half of the paper, we discuss the energy reflected by a moving perfect mirror and point out
that the formula in Einstein's 1905 paper is correct only in two limiting cases.
There has been some confusion in the Breakthrough Starshot literature, in part due to Einstein's
failure to point out the assumptions in his derivation, that we have cleared up.

%\section{Author Declarations}
%The authors have no conflicts to disclose.

%\clearpage

%\bibliography{paper}


\end{document}

\title{Supplementary Information: Energy Needed to Propel a Tiny Spacecraft to Proxima Centauri,
and,
An Unstated Assumption in Einstein's 1905 Paper}

\author{C. J. Umrigar\footnote{CyrusUmrigar@cornell.edu}}
\affiliation{
Laboratory of Atomic and Solid State Physics,\\
Cornell University, Ithaca, NY 14853, USA.}

\author{Tyler A. Anderson\footnote{taa65@cornell.edu}}
\affiliation{
Laboratory of Atomic and Solid State Physics,\\
Cornell University, Ithaca, NY 14853, USA.}

%\maketitle
{
\let\clearpage\relax
\maketitle
}

\section{Exact Formula for the Time Needed to Reach $\beta$}

In derivation 2 of the paper we find a formula for the total energy emitted by the laser at the instant the craft reaches a particular $\beta$.
Dividing this expression by the power $P$ of the laser we obtain an expression for the time in the Earth frame when the craft reaches a speed $\beta$:
\beq
t &=& {m_0 c^2 \over 2 P } \left({(2-\beta)\sqrt{1-\beta^2} \over 3(1-\beta)^2} -{2 \over 3} \right).
\label{eq:t_continuum}
\eeq
%where
%\beq
%t_{\rm rel} = {m_0 c^2 \over P}.
%\eeq
This formula is strictly correct only in the continuum (classical electrodynamics) limit where the discrete nature of the photons emitted by the laser is ignored.
We now derive an exact expression which is correct even outside the continuum limit and show that the correction to Eq.~\ref{eq:t_continuum} is extremely small - on the order of $10^{-33}$ for Breakthrough Starshot.
Although the beginning of our analysis resembles one already given by Kipping in Ref.~\onlinecite{Kip-AJ-17-err}, we make no approximations and arrive at a very different conclusion.

Suppose the laser emits photons of energy $E_{\gamma} = r m_0 c^2$ (recall that $m_0$ is the mass of the craft) at intervals of $T$ so that $P = {r m_0 c^2 \over T}$.
Let $t_n$ be the time when $n$ photons have been received by the craft and let $\beta_n$ be the speed of the craft immediately after the $n$th photon has been received.
We have already shown in the paper that, for a craft moving at $\beta$, the time between consecutive photon collisions with the sail is just
\beq
T_r = { T \over 1 - \beta},
\eeq
so that we may find the cumulative time $t_n$ by summing over all these times:
\beq
t_n = \sum^{n-1}_{m=0} {T \over 1 - \beta_{m}}.
\label{eq:tn}
\eeq
The total energy incident on the craft immediately after $t_n$ is just
\beq
E_n = n E_{\gamma} = n r m_0 c^2.
\label{eq:En}
\eeq
Using the formula from derivation 1 of the paper, we can relate the total energy incident on the craft $E_n$ (and thus $n$) to $\beta_n$:
\beq
nr = {E_{n} \over m_0 c^2}  = {1 \over 2} \left( \sqrt{1 + \beta_n \over 1 - \beta_n} - 1 \right).
\label{eq:nr}
\eeq
This expression can then be inverted to find $\beta_n$ as a function of $n$:
%Combining Eqs.~\ref{eq:En1} and \ref{eq:En2} and solving for $\beta_n$ in terms of $n$ gives
\beq
\beta_n = {(2nr + 1)^2 -1 \over (2nr + 1)^2 + 1},
\eeq
so that
\beq
{1 \over 1 - \beta_n} = {  (2nr + 1)^2 + 1 \over 2}.
\eeq
Substituting this expression into Eq.~\ref{eq:tn} and using the elementary formulae
$$\sum^{n-1}_{m=0} 1 = n, $$
$$\sum^{n-1}_{m=0} m = {n(n-1) \over 2}, $$
$$\sum^{n-1}_{m=0} m^2 = {n(n-1)(2n-1) \over 6}$$
gives
\beq
t_n &=& T \sum^{n-1}_{m=0} \left(1 + 2 n r + 2 (nr)^2\right) \nonumber \\
    &=& T \left( n + r n (n-1) + {1 \over 3} r^2 n (n-1) (2n-1) \right).
\eeq
Rearranging and using ${T \over r} = {m_0 c^2 \over P} $ we find
\beq
t_n = {m_0 c^2 \over P} \left(  nr + (nr)^2 + {2 \over 3} (nr)^3 - nr^2 - n^2 r^3 + {1 \over 3} n r^3 \right). 
\label{eq:tn2}
\eeq
%In the continuum limit the energy per photon decreases to zero so that $r \rightarrow 0$ but the total number of photons increases to infinity $n \rightarrow \infty$ in such a way that the product $nr$ is finite.
In the continuum limit $r \rightarrow 0$ but $n \rightarrow \infty$ in such a way that the product $nr$ remains finite.
Thus, we expect the first three terms in Eq.~\ref{eq:tn2} to agree with Eq.~\ref{eq:t_continuum} while the remaining three terms represent a correction that goes to zero in the continuum limit.
We now show that this is indeed the case by using Eq.~\ref{eq:nr} to write these three terms in terms of $\beta_n$:
\beq
nr = {\color{red} {1 \over 2} \sqrt{1 + \beta_n \over 1 - \beta_n}} - {1 \over 2}
\eeq
\beq
(nr)^2  = {\color{blue} {1 \over 4} \left( {1 + \beta_n \over 1 - \beta_n} \right) }  {\color{red} - {1 \over 2} \sqrt{1 + \beta_n \over 1 - \beta_n}} + {1 \over 4}
\eeq
\beq
{2 \over 3} (nr)^3  = {1 \over 12} \left(\sqrt{1 + \beta_n \over 1 - \beta_n}\right)^3 {\color{blue} - {1 \over 4} \left({1 + \beta_n \over 1 - \beta_n}\right) }  + {1 \over 4} \sqrt{1 + \beta_n \over 1 - \beta_n}  - {1 \over 12}.
\eeq
Adding these three equations together we see that there are a number of immediate cancellations (indicated by terms of the same color), and we are left with
\beq
nr + (nr)^2 + {2 \over 3} (nr)^3 
&=& {1 \over 12} \left(\sqrt{1 + \beta_n \over 1 - \beta_n}\right)^3
+ {1 \over 4} \sqrt{1 + \beta_n \over 1 - \beta_n}
- {1 \over 3} \\
&=& {1 \over 6} {(2-\beta_n)\sqrt{1 - \beta^2_n} \over (1 - \beta_n)^2 } - {1 \over 3}.
\eeq
Substituting the above expression into Eq.~\ref{eq:tn2} we finally obtain
\beq
t_n = {m_0 c^2 \over 2 P} \left( {(2-\beta_n)\sqrt{1 - \beta^2_n} \over 3 (1 - \beta_n)^2 } - {2 \over 3} \right) + \delta t_n %- nr^2 - n^2 r^3 + {1 \over 3} n r^3 \right). 
\eeq
where
\beq
%\delta t_n = - {m_0 c^2 \over P} \left( nr^2 + n^2 r^3 - {1 \over 3} n r^3 \right). \\
\delta t_n = - {m_0 c^2 \, nr \over P} \left( r + \left(n-{1 \over 3}\right) r^2 \right).
\eeq
%Note that this correction is negative, in agreement with Kipping's observation that the nonzero recoil of the photons causes the craft to reach a given speed at an earlier time.
Note that this correction is negative, but that is because the times $t_n$ that we have chosen
are immediately after the $n$th photon has struck.
If instead, we had defined $t_n$ to be the time just before the $n$th photon has struck, then the
opposite sign would be obtained.
%Note that this correction is always negative.  This can be understood as follows.  First note that immediately
%after $n$ photons worth of energy have struck the sail, the velocity is the same in the continuous
%and discrete cases.  In the continuous case, the spacecraft starts to move at $t=0$, whereas in the
%discrete case it starts to move when the first photon hits the sail at time $T$.
%Hence, at any time, the continous-case spacecraft is further from the laser than the discrete-case spacecraft.
%This means that when the $n$th photon hits the spacecraft in the discrete case, not all of the $n$ photons
%worth of energy has reached the spacecraft in the continuous case.  Hence, at any time later than $T$, the
%spacecraft has a slightly lower velocity in the continuous case than in the discrete case,
%or equivalently, the time needed to reach velocity $\beta$ is slightly smaller in the
%discrete case than in the continuous case.
%However, we also see that because
%Note that
%$nr$ is ${\cal O}(1)$ and $r$ is ${\cal O}(10^{-33})$, $\delta t_n$ is ${\cal O}(10^{-33})$ which is
%totally negligible.
What is important is that since
$nr$ is ${\cal O}(1)$ and $r$ is ${\cal O}(10^{-33})$, $\delta t_n$ is ${\cal O}(10^{-33})$ which is
totally negligible.

%\clearpage

%\bibliography{paper}